\documentclass[epj]{webofc}
\usepackage[varg]{txfonts}
\usepackage[utf8]{inputenc} 
\usepackage[usenames]{color}
\usepackage{comment}
\usepackage{amsmath}
\usepackage[utf8]{inputenc}

\def\beq{\begin{equation}}
\def\eeq{\end{equation}}
\def\bea{\begin{eqnarray}}
\def\eea{\end{eqnarray}}
\def\D0{D\O }

\wocname{EPJ Web of Conferences}
\woctitle{ICNFP 2016}
\begin{document}
\selectlanguage{english}
\title{Semileptonic decays and $|V_{xb}|$ determinations}
%
%

\author{Giulia Ricciardi\inst{1,2}\fnsep\thanks{\email{giulia.ricciardi@na.infn.it}}}

\institute{Dipartimento di Fisica E. Pancini, Universit\`a  di Napoli Federico II,
Complesso Universitario di Monte Sant'Angelo, Via Cintia,
I-80126 Napoli, Italy
\and
INFN, Sezione di Napoli,
Complesso Universitario di Monte Sant'Angelo, Via Cintia,
I-80126 Napoli, Italy
}

\abstract{%
 We briefly summarize  up-to-date results on the determination of the  parameters of the Cabibbo-Kobayashi-Maskawa  matrix  $|V_{cb}|$ and $|V_{ub}|$, which
play an important role in the unitarity triangle and in testing the Standard Model, and  recent results on semileptonic $B$ meson decays involving a $\tau$ lepton.}
\maketitle
\section{Introduction}
\label{intro}
 We briefly review recent results on the semi-leptonic B decays  and on the determination of the  parameters of the Cabibbo-Kobayashi-Maskawa (CKM) matrix  $|V_{cb}|$ and $|V_{ub}|$, which
play an important role in the unitarity triangle and in testing the Standard Model (SM). For instance, the parameter $\epsilon_K$ depends on $|V_{cb}|^4$, while the ratio $|V_{ub}/V_{cb}|$ directly constrains one side of the unitarity triangle.  The SM does not predict the values of the CKM matrix elements and the most precise measurements of $|V_{cb}|$ and $|V_{ub}|$ come from semi-leptonic decays, that being tree level at the lowest order in the SM are generally considered unaffected by new physics.
The   inclusive and exclusive semi-leptonic  searches  rely on
different theoretical calculations  and on
different experimental techniques which have, to a large extent, uncorrelated
statistical and systematic uncertainties. This independence makes the agreement between determinations of
 $|V_{cb}|$ and $|V_{ub}|$ values from inclusive and exclusive decays a useful test of our understanding of experimental data extraction and underlying theory (see e.g. \cite{Ricciardi:2013xaa,Ricciardi:2014aya,Ricciardi:2014iga,Ricciardi:2016pmh,  Koppenburg:2017mad} and references therein). We  discuss up-to-date tensions  between the  inclusive and exclusive determinations of
 $|V_{cb}|$ and $|V_{ub}|$ within the SM and  recent results on semileptonic $B$ meson decays involving a $\tau$ lepton.

\section{Exclusive $|V_{cb}|$ determination}
\label{subsectionExclusive decays}

For negligible lepton masses ($\ell=e, \mu)$,
the  differential ratios for the semi-leptonic CKM favoured decays $B \to D^{(\ast)} \ell \nu$
can be written as
\begin{eqnarray}
&\frac{d\Gamma}{d \omega}(B \rightarrow D^\ast\,\ell {\nu})&
\propto G_F^2    (\omega^2-1)^{\frac{1}{2}}
 |V_{cb}|^2   {\cal F}(\omega)^2
 \nonumber \\
 &\frac{d\Gamma}{d \omega} (B \rightarrow D\,\ell {\nu})&  \propto
G_F^2\,
(\omega^2-1)^{\frac{3}{2}}\,
 |V_{cb}|^2   {\cal G}(\omega)^2
 \label{diffrat}
\end{eqnarray}
The  recoil parameter $\omega = p_B \cdot p_{D^{(\ast)}}/m_B \, m_{D^{(\ast)}}$ corresponds to the energy transferred to the leptonic pair. For the exact expression of the differentials in Eq. \eqref{diffrat} we refer  to the current literature. Here we care to emphasize the dependence on a single form factor, ${\cal F}(\omega)$  for $B \to D^{\ast} l  \nu$ and   ${\cal G}(\omega)$ for $B \to D l  \nu$,  and the phase space vanishing at the  no-recoil point $\omega=1$ in both cases.

 In the
heavy quark limit
both form factors are related to a single Isgur-Wise
function,  ${\cal F(\omega) }= {\cal G(\omega) } = {\cal  \xi (\omega) }  $, which is
normalized to unity at zero recoil,  that is  ${\cal \xi (\omega}=1) =1 $.
There are
non-perturbative corrections  to this prediction, expressed at the zero-recoil point by the heavy quark symmetry  under the form of
powers of $\Lambda_{QCD}/m$, where $m= m_c$ and $m_b$. Other corrections are perturbatively calculable radiative
corrections from hard gluons and photons.

In order to extract the CKM factors, we  need  not only to compute the form factors, but also to measure experimental decay rates, which
  vanish at zero-recoil.
Therefore,  experimental points are extrapolated to zero recoil, using a parametrization of the
dependence on $\omega$ of the form factor.

Recent determinations adopt  a  parametrization where
 $\omega$ is mapped onto a complex variable $z$ via the conformal transformation
$
z= (\sqrt{\omega+1}-\sqrt{2})/(\sqrt{\omega+1}+\sqrt{2})
$. The form factors may be written in form of an
expansion in $z$, which converges rapidly in the kinematical
region of heavy hadron decays. The coefficients of the expansions  are subject to unitarity  bounds  based  on  analyticity.
 Common examples  are the
 CLN
(Caprini-Lellouch-Neubert) \cite{Caprini:1997mu},
 the BGL
(Boyd-Grinstein-Lebed)  \cite{Boyd:1994tt} and the  BCL (Bourrely-Caprini-Lellouch) \cite{Bourrely:2008za} parameterizations.
They are  all  constructed  to  satisfy  the  unitarity  bounds, but the CLN approach differs
mostly  in its reliance  on  next-to-leading  order
HQET  relations  between  the  form  factors.
Recently, the reliability of the CLN approach has been questioned in both $B \to D\, l \nu  $ \cite{Bigi:2016mdz} and $B \to D^\ast \, l \nu$ \cite{Bigi:2017njr, Grinstein:2017nlq} channels.

The experiments, by measuring the differential decay rates with  a variety of methods, provide inputs for several fits, that, among other parameters, aim at estimating the CKM values. A combined  fit  of  the
$B \to D^{(\ast)} \, \ell \nu$
differential  rates
and angular distributions, consistently including the HQET relations to
$O(\Lambda_{QDE}/m_{c,b}, \alpha_s)$, has recently been performed.
Under various fit scenarios, that use or omit lattice QCD and QCD sum rule predictions, they
constrain the leading and subleading Isgur-Wise functions \cite{Bernlochner:2017jka}.

\subsection{$B \to D^{\ast} \ell  \nu$ channel}

Until now, the FNAL/MILC  collaboration  has been the only one performing
the  non perturbative determination  of the form factor ${\cal F}(1)$, at zero recoil,  for the $B \to D^{\ast} \ell  \nu$ channel
in the lattice unquenched $N_f= 2+1$  approximation, and  their latest estimate gives  the value \cite{Bailey:2014tva}
\beq  {\cal F}(1)
=0.906\pm 0.004 \pm  0.012  \label{VcbexpF2}  \eeq
The first error is statistical and the second one  is the sum in quadrature of all systematic errors.
The lattice QCD theoretical error is now commensurate with the experimental error (they contribute respectively for about 1.4\% and 1.3\%), while  the QED error contributes for about  0.5\%.
Large discretization error could be in principle be reduced  by going to finer lattice spacings
or larger lattice sizes.
The total uncertainty is around  the (1-2)\% level.

There are two recent $|V_{cb}|$ determinations from the Heavy Flavour and Lattice Averaging Groups, HFLAV and FLAG respectively, that use the form factor \eqref{VcbexpF2}; we report them in  Table \ref{phidectab2}.
 Using  the CLN parametrization,
the 2016 HFLAV average \cite{Amhis:2016xyh}
gives
\beq |V_{cb}| = (39.05 \pm 0.47_{\mathrm{exp}} \pm 0.58_{\mathrm{th}} ) \, \mathrm{x} \, 10^{-3} \label{ll1} \eeq
where the first uncertainty is experimental and the second error is theoretical (lattice QCD
calculation and electro-weak correction).
The 2016 FLAG  $N_{f}=2+1$ $|V_{cb}|$  average value yields \cite{Aoki:2016frl}
\beq |V_{cb}| = (39.27 \pm 0.49_{\mathrm{exp}} \pm 0.56_{\mathrm{latt}}) \times\, 10^{-3} \label{ll1flag} \eeq
This average employs  the 2014 HFLAV experimental average  \cite{Amhis:2014hma} ${\cal F}(1) \eta_{\mathrm{EW}} |V_{cb}|= (35.81 \pm 0.45) \times\, 10^{-3}$ and  the value $\eta_{\mathrm{EW}} = 1.00662$.

The HPQCD collaboration has presented preliminary results for the $B \to D^\ast$ form factor at zero recoil,
based on
relativistic HISQ charm quark and NRQCD bottom quark,
giving the estimate
 $|V_{ub}|= (41.5 \pm 1.7 )\times 10^{-3}$ \cite{Harrison:2016gup}.


Many experiments  have measured the differential decay rate as a function
of $\omega$, but only
recently, and for the first time, the unfolded
fully-differential decay rate and associated covariance matrix have been published, by the Belle collaboration  \cite{Abdesselam:2017kjf}. Using the CLN parametrization and the lattice form factor value, they extract the value \cite{Abdesselam:2017kjf}
\beq
|V_{cb}| = (37.04 \pm 1.3) \, \times\, 10^{-3}
\eeq
Using Belle data, it has been  shown that   when switching from
the CLN to the BGL form the determination of $|V_{cb}|$ shifts  beyond the quoted experimental precision \cite{Grinstein:2017nlq,Bigi:2017njr}. These analyses
are consistent with each other and give in the BGL framework, along with the lattice value given for the zero recoil form factor, the values
\cite{Grinstein:2017nlq}
 \beq
 |V_{cb}| = (41.9^{+2.0}_{-1.9}) \, \times \, 10^{-3} \eeq
 and \cite{Bigi:2017njr}
\beq |V_{cb}| = (41.7^{+2.0}_{-2.1}) \, \times\, 10^{-3}
\eeq
The central value is higher than the corresponding value in CLN parametrization.
However, it has also been argued that
 fits that yield the higher values of
$|V_{cb}|$
 suggest large violations of heavy quark symmetry
and   tension  with   lattice  predictions of  the  form factor ratios \cite{Bernlochner:2017xyx}.

Moving to estimates of the
 form factor  via zero recoil sum rules, we have
 \cite{Gambino:2010bp, Gambino:2012rd}
\beq {\cal F}(1) = 0.86 \pm 0.01 \pm 0.02 \label{gmu} \eeq
where the second uncertainty accounts for the
excited states.
This value is in good agreement with the lattice value in Eq. \eqref{VcbexpF2}, but  slightly lower in the central value. That implies  a relatively higher value of $|V_{cb}|$, that is
\beq  |V_{cb}| = (41.6\pm 0.6_{\mathrm{exp}}\pm 1.9_{\mathrm{th}})  \, \mathrm{x} \, 10^{-3}
\label{wee}
\eeq
where the HFAG averages \cite{Amhis:2012bh} have been used. The theoretical error  is more than twice the error in the lattice determination \eqref{ll1}.

\subsection{The $B \to D \ell  \nu$ channel}

For $ B \rightarrow D \, \ell \, \nu$ decay,
 the FNAL/MILCcollaboration has calculated in 2015 the form factors in the unquenched lattice-QCD approximation \cite{Lattice:2015rga}
for a range of recoil momenta. By parameterizing their dependence on momentum transfer
using  the BGL z-expansion, they determine
$ |V_{cb}| $
from the relative normalization over the entire range of recoil momenta, which reads
 \cite{Lattice:2015rga}
\beq
 |V_{cb}| =(39.6 \pm 1.7_{\mathrm{exp+QCD}} \pm 0.2_{\mathrm{QED}})   \, \mathrm{x} \, 10^{-3}
\eeq
The   average value is almost the same than the  one inferred from
$ B \rightarrow D^\ast \, \ell \, \nu$ decay by the same collaboration, see Eq. \eqref{ll1} and  Table \ref{phidectab2}.

Results on
$ B \rightarrow D\, \ell \, \nu$
form factors at non-zero recoil have  also been given the same year by the HPQCD Collaboration \cite{Na:2015kha}. Their results are based    on the non-relativistic QCD (NRQCD) action
for  bottom  and  the  Highly  Improved  Staggered  Quark
(HISQ) action for charm quarks, together with $N_f=2+1$ MILC gauge configuration.
A  joint  fit to  lattice  and 2009  BaBar  experimental  data  \cite{Aubert:2009ac} allows the
extraction of the CKM matrix element  $|V_{cb}| $, using the CLN parametrization. It gives \cite{Na:2015kha}
\beq
|V_{cb}|= (40.2 \pm 1.7_{\mathrm{latt+stat}} \pm 1.3_{\mathrm{syst}})   \, \mathrm{x} \, 10^{-3}
\eeq
The  first  error  consists  of  the  lattice  simulation  errors  and  the  experimental  statistical  error  and
the  second  error  is  the  experimental  systematic  error.
The dominant error  is the discretization error, followed by higher order current matching uncertainties.  The former error can be reduced by adding
simulation  data  from  further  ensembles  with  finer  lattice spacings.

%

In  2015 the decay $ B \rightarrow D\, \ell \, \nu$ has also been measured in fully reconstructed events  by the Belle collaboration \cite{Glattauer:2015teq}, They have  performed a fit to the CLN parametrization, which has two free parameters, the form factor at zero recoil ${\cal G}(1)$ and the linear slope $\rho^2$.
The fit has been used to determine $\eta_{EW} {\cal G}(1) |V_{cb}|$, that, divided by the form-factor normalization ${\cal G}(1)$ found by the FNAL/MILC Collaboration
\cite{Lattice:2015rga}, gives
$
\eta_{EW}  |V_{cb}|=(40.12 \pm 1.34) \times 10^{-3}
$  \cite{Glattauer:2015teq}.
Assuming $\eta_{EW}   \simeq 1.0066 $, it  translates into  \cite{Glattauer:2015teq}\beq
 |V_{cb}|=(39.86 \pm 1.33) \times 10^{-3}\eeq The Belle Collaboration also obtain
a slightly more precise
result (2.8\% vs.  3.3\%)  by
 exploiting  lattice data at non-zero recoil and  performing   a combined fit to the BGL form factor. It yields
$
\eta_{EW}  |V_{cb}|=(41.10 \pm 1.14) \times 10^{-3}
$
which translates into \cite{Glattauer:2015teq} \beq
 |V_{cb}|=(40.83 \pm 1.13) \times 10^{-3}
\eeq
assuming once again $\eta_{EW}   \simeq 1.0066 $.

The latest lattice results, as well as \cite{Lattice:2015rga, Na:2015kha}, Belle \cite{Glattauer:2015teq} and Babar \cite{Aubert:2009ac} data, have been used in a global fit in the BGL parametrization which gives,
in agreement with previous results \cite{Bigi:2016mdz} \beq
 |V_{cb}|=(40.49 \pm 0.97) \times 10^{-3} \eeq
In \cite{Bigi:2016mdz} differences on  BGL, CLN, and BCL parameterizations are discussed.

\section{Inclusive $|V_{cb}|$ determination}
\label{subsectionInclusive decays}

In  inclusive $ B \rightarrow X_c \, \ell \, \nu_l$ decays,  the final state
$X_c$ is an hadronic state originated by the charm  quark. There is no dependence on the details of the final state, and quark-hadron duality is generally assumed.
Sufficiently inclusive quantities (typically the width
and the first few moments of kinematic distributions) can be expressed as a double series in $\alpha_s$ and $\Lambda_{QCD}/m$, in the framework of   the Heavy Quark Expansion (HQE),  schematically indicated as
\begin{equation}
\Gamma(B\rightarrow X_c l \nu)=\frac{G_F^2m_b^5}{192 \pi^3}
|V_{cb}|^2 \left[ c_3 \langle O_3 \rangle +
c_5\frac{ \langle O_5 \rangle }{m_b^2}+c_6\frac{ \langle O_6 \rangle }{m_b^3}+O\left(\frac{\Lambda^4_{QCD}}{m_b^4},\; \frac{\Lambda^5_{QCD}}{m_b^3\, m_c^2}, \dots \right)
\right] \label{HQE}
\end{equation}
Here  $c_d$ ($d=3,5,6 \dots$) are short distance coefficients, calculable  in perturbation theory as a series in the strong coupling $\alpha_s$, and
$O_d$ denote local operators of (scale) dimension $d$.
The hadronic
expectation values of the operators  $\langle O_d \rangle $ encode the
nonperturbative corrections and can be parameterized in terms of  HQE  parameters,
whose number grows with
powers of $\Lambda_{QCD}/m_b$.
Similar expansions give the moments of distributions of
charged-lepton energy, hadronic invariant mass and hadronic energy.

Let us observe that the first order in the series corresponds to the parton order, while  terms of order $\Lambda_{QCD}/m_b$ are absent.
At order $1/m_b^0$ in the HQE, that is at the parton level,  the  perturbative corrections up to order $\alpha_s^2$ to the width and to the moments of the lepton energy and hadronic mass
distributions are known completely (see Refs. \cite{Trott:2004xc, Aquila:2005hq, Pak:2008qt, Pak:2008cp, Biswas:2009rb}
and references therein). The terms of order $\alpha_s^{n+1} \beta_0^n$, where $\beta_0$ is the first coefficient of the QCD $\beta$ function, $\beta_0= (33-2 n_f)/3$, have also been computed following  the
 Brodsky-Lepage-Mackenzie (BLM) procedure \cite{Aquila:2005hq, Benson:2003kp}.

The next order  is $ \Lambda_{QCD}^2/m_b^2$, and at this order the HQE includes
two  operators, called the kinetic energy  and the chromomagnetic operator,  $\mu^2_{\pi}$ and  $\mu^2_{G}$.
Perturbative corrections to the coefficients   of the kinetic operator  \cite{Becher:2007tk,Alberti:2012dn}
and  the chromomagnetic operator
\cite{Alberti:2013kxa, Mannel:2014xza, Mannel:2015jka}   have been
 evaluated   at order $\alpha_s^2$ .

Neglecting  perturbative corrections, i.e. working at tree level,  contributions to various observables   have been
computed at order $1/m_b^3$ \cite{Gremm:1996df} and estimated at order $1/m_b^{4,5}$  \cite{Dassinger:2006md, Mannel:2010wj,Heinonen:2014dxa}.

Starting at  order  $\Lambda_{QCD}/m_b^3$,   terms
 with an infrared sensitivity to the charm mass,
 appear, at this order as a $\log m_c$ contribution \cite{Bigi:2005bh, Breidenbach:2008ua, Bigi:2009ym}. At higher orders these  contributions, sometimes dubbed intrinsic charm  contribution,
in form of powers of
 $\Lambda_{QCD}/m_c$ have to be considered as well.  Indeed, roughly speaking, since $m^2_c \sim O( m_b \Lambda_{\mathrm{QCD}})$ and $\alpha_s(m_c) \sim O(\Lambda_{\mathrm{QCD}})$, contributions of order
 $\Lambda^5_{\mathrm{QCD}}/m^3_b \, m^2_c$
and $\alpha_s(m_c) \Lambda^4_{\mathrm{QCD}}/m^2_b\, m^2_c
$  are expected
comparable in size to  contributions of order $\Lambda^4_{\mathrm{QCD}}/m^4_b$.
The HQE parameters are affected by the
 particular theoretical framework (scheme) that is
used to define the quark masses.

In HQE the number of  nonperturbative parameters grows with the order in $1/m_b$. At leading order, the matrix elements can  be reduced to one, while at dimension-four heavy-quark symmetries and the equations of motion
ensure that the forward matrix elements of the operators can be expressed in terms of the
matrix elements of higher dimensional operators. The first nontrivial contributions appear at
dimension five, where two independent parameters, $\mu^2_{\pi,G}$, are needed, and two independent parameters, $\rho^3_{D,LS}$, are also needed at dimension six.
At dimension seven  and eight, nine and  eighteen independent matrix elements appear, respectively, and for higher orders one has an almost
factorial increase of the number of independent parameters.
These parameters depend on the heavy quark mass, although sometimes the infinite mass
limits of these parameters is taken.

The rates and the spectra are very sensitive to $m_b$.
 The physical pole mass definition for heavy quark masses is not a reasonable choice, because of problems in the convergence of
 perturbative series for the decay rates \cite{Olive:2016xmw,Hoang:2017btd}. Other possibilities are
 the use of “short-distance” mass definitions, such as the kinetic
scheme \cite{Bigi:1994ga}, the 1S scheme \cite{Hoang:1998hm}, or
 the $\overline MS$ mass, $m_b^{\overline MS}(m_b)$.
The 1S scheme eliminates the
b quark pole mass by relating it to the perturbative expression for the mass of the 1S state
of the $\Upsilon$ system. In the kinetic scheme, the so-called “kinetic mass” $m^{kin}_b(\mu)$
is the mass entering the non-relativistic expression for the kinetic energy of a heavy
quark, and is defined using heavy-quark sum rules.
The alternative are short-distance mass definitions, as the   $\overline{MS}$ masses. However, the scale $m_b$ for  $m_b^{\overline{MS}}(m_b)$ is generally considered
  unnaturally high for B decays, while  $m_b^{\overline{MS}}(\mu)$ at smaller scales ($\mu \sim 1$ GeV) is under
poor control.

A global fit   is a simultaneous fit to
 HQE  parameters, quark masses and absolute values of  CKM matrix elements obtained by  measuring
 spectra plus all
available moments.
The semileptonic moments alone determine only a linear combination of $m_b$
and $m_c$, and
additional input is required to allow a precise determination of $m_b$. This additional
information can come from the radiative $B \to X_s \gamma$ moments or from precise determinations of the
charm quark mass.
The
HFLAV global fit \cite{Amhis:2016xyh}  employs as  experimental inputs  the (truncated) moments of the
lepton energy $E_l^n$  (in the $B$ rest frame) and the $m_X^n$ momenta in the hadron spectra in $B \to X_c \ell \nu$.
It is performed in the kinetic scheme, includes 6  non-perturbative  parameters ($m_{b,c}$, $\mu^2_{\pi,G}$,  $\rho^3_{D,LS}$) and the charm mass as  the additional constraint,   yielding
  \beq |V_{cb}| = (42.19\pm 0.78) \times 10^{-3} \label{HFAGincl16}\eeq
In the same kinetic scheme,  another global fit, including the complete power corrections up to $O(\alpha_s\Lambda_{QCD}^2/m_b^2)$, has been performed, giving the estimate $|V_{cb}| = (42.21 \pm 0.78) \times 10^{-3}$ \cite{Alberti:2014yda}. More recently, the effect of including  $1/m_b^{4, 5}$ corrections in the global fit has  been also analyzed, in the so-called Lowest-Lying State Approximation (LLSA), which assumes that the lowest lying heavy meson states
saturate a sum-rule for the insertion of a heavy meson
state sum \cite{Mannel:2010wj, Heinonen:2014dxa, Gambino:2016jkc}. The  LLSA  was used because of the large number of new parameters, in order to provide
loose constraints on the higher power matrix elements.
A resulting  global fit to the semileptonic moments  in  the LLSA gives the estimate  \cite{Gambino:2016jkc}
\beq
|V_{cb}| = (42.11 \pm 0.74) \times 10^{-3}
\label{gambincl16}
\eeq
Indirect $|V_{cb}|$ estimates from CKMfitter \cite{CKMfitter}, using a frequentist statistical approach,  and UTfit \cite{Utfit16} Collaborations, adopting instead a Bayesian
approach, are reported in Table \ref{phidectab2}.

Let us mention that this year a method to non-perturbatively calculate the forward-scattering matrix elements
relevant to inclusive semi-leptonic
$B$
meson decays on lattice
has been proposed \cite{Hashimoto:2017wqo}.

\begin{table}[h]
\centering
\caption{Status of exclusive  and  inclusive $|V_{cb}|$  determinations}
\label{phidectab2}
\begin{tabular}{lrr}
\hline
 \hline
{ \color{red}{Exclusive decays}}
& {\color{red}{ $ |V_{cb}| \times  10^{3}$}}
  \\
\hline
{\color{blue}{ $\bar{B}\rightarrow D^\ast \, l \, \bar{\nu}$ }}  & \\
\hline
Grinstein et al. 2017  (Belle data, BGL) \cite{Grinstein:2017nlq} & $ 41.9^{+2.0}_{-1.9}$ \\
Bigi et al. 2017 (Belle data, BGL) \cite{Bigi:2017njr} & $ 41.7^{+2.0}_{-2.1}$ \\
Belle 2017 (CLN) \cite{Abdesselam:2017kjf} & $ 37.04 \pm 1.3$ \\
FLAG 2016 \cite{Aoki:2016frl} & $ 39.27 \pm 0.49_{\mathrm{exp}} \pm 0.56_{\mathrm{latt}} $ \\
HFLAV 2016 (FNAL/MILC 2014 $\omega=1$) \cite{Amhis:2016xyh}   & $ 39.05 \pm 0.47_{\mathrm{exp}} \pm 0.58_{\mathrm{th}}  $ \\
HFAG 2012 (Sum Rules) \cite{ Gambino:2010bp, Gambino:2012rd, Amhis:2012bh} & $   41.6\pm 0.6_{\mathrm{exp}}\pm 1.9_{\mathrm{th}} $ \\
\hline
{\color{blue}{ $  \bar{B}\rightarrow D \, l \, \bar{\nu} $ }} &   \\
\hline
Global fit   2016 \cite{Bigi:2016mdz}  &  $40.49 \pm 0.97$
\\
Belle 2015 (CLN)   \cite{Glattauer:2015teq,Lattice:2015rga}  & $ 39.86 \pm 1.33 $
 \\
Belle 2015 (BGL)   \cite{Glattauer:2015teq, Lattice:2015rga, Na:2015kha}  & $40.83 \pm 1.13$
 \\
FNAL/MILC  2015 (Lattice  $\omega \neq 1)$ \cite{Lattice:2015rga}  & $39.6 \pm 1.7_{\mathrm{exp+QCD}} \pm 0.2_{\mathrm{QED}} $\\
HPQCD  2015 (Lattice $\omega \neq 1)$  \cite{Na:2015kha}  & $40.2 \pm 1.7_{\mathrm{latt+stat}} \pm 1.3_{\mathrm{syst}}
$
 \\
 \hline
{ \color{red}{Inclusive decays}}
  \\
\hline
HFLAV 2016  \cite{Amhis:2016xyh} & $ 42.19 \pm 0.78 $ \\
 Gambino et al. 2016 \cite{Gambino:2016jkc}  & $42.11 \pm 0.74$ \\
\hline
{ \color{red}{Indirect fits}}
\\
\hline
UTfit  2017 \cite{Utfit16} &
$ 42.7 \pm  0.7 $
\\
CKMfitter  2016 ($3 \sigma$) \cite{CKMfitter} &
$41.81^{+0.91}_{-1.81}$
\\
\hline
\hline
\end{tabular}
\end{table}


\section{Exclusive $|V_{ub}|$ determination}

The parameter $|V_{ub}|$ is the less precisely known among the modules of the CKM matrix elements.
The CKM-suppressed decay $B \to \pi \ell \nu$ with light final leptons is the typical exclusive channel used to extract $|V_{ub}|$.
It is well-controlled experimentally and several measurements have been performed by both
 BaBar and Belle collaborations \cite{Hokuue:2006nr, Aubert:2006ry, Aubert:2008bf, delAmoSanchez:2010af, Ha:2010rf, Lees:2012vv, Sibidanov:2013rkk}.

Commonly used non-perturbative approaches to form factor calculations are lattice QCD (LQCD) and light-cone sum rules (LCSR).
At low $q^2$,
i.e.
when the mass of the
B-meson must be balanced by a large pion momentum in order to transfer a small
momentum to the lepton pair, lattice computations present large discretization errors and very
large statistical errors. The high
$q^2$
region is much more accessible to the lattice. On the other side, the low $q^2$ region is the range of applicability of LCSR.

The   lattice determinations of  $f_+(q^2)$  in the $B \to \pi l \nu$ channel, based on unquenched  simulations, have been obtained by  the HPQCD
\cite{Dalgic:2006dt}, the Fermilab/MILC \cite{Bailey:2008wp, Lattice:2015tia} and the RBC/UKQCD \cite{Flynn:2015mha}  collaborations.
The Fermilab/MILC collaboration has evaluated the form factor $f_+(q^2=20 \, {\mathrm{GeV}})$ with an uncertainty going down to 3.4\%. Leading contribution to the uncertainty  come from the chiral-continuum extrapolation fit, including statistical
and heavy-quark discretization errors.
%
%

%

In 2016 the HPQCD collaboration has presented 2+1+1-flavor results for
$ B \to \pi \ell  \nu$ decay at zero recoil, with the $u/d$ quark
masses going down to their physical values, for the first time; they also calculated the scalar factor $f_0$ form at zero recoil to 3\% precision
\cite{Colquhoun:2015mfa}.

At
large recoil (small $q^2$),
 direct LCSR calculations of the semi-leptonic  form factors  are available, which have benefited by   progress in pion distribution amplitudes, next-to-leading and leading  higher order twists and QCD corrections (see e.g. Refs.~\cite{Khodjamirian:2011ub, Bharucha:2012wy,Li:2012gr,Imsong:2014oqa, Wang:2015vgv} and references within).
%
%

Branching  fraction  measurements  of  semileptonic  $B$ decays
 are  possible  using
several different experimental techniques that differ  in the way  the companion
$B$
meson
is reconstructed. In untagged analyses, the signal
$B$
meson is reconstructed, with the exception of the
escaped neutrino.  The 4-momentum of the companion
$B$ meson is inclusively determined
 by adding
up the 4-momenta of all the remaining charged tracks and neutral clusters in the event.
Since the initial  state $\Upsilon(4S)$ is well-known,  the missing 4-momentum can be identified with the neutrino 4-momentum,  if neutrino is the only missing particle
in the event. In tagged analyses, the companion $B$ meson is fully reconstructed in either a semileptonic or an hadronic way.
The available state-of-the-art experimental input consists
of  three untagged measurements by BaBar \cite{delAmoSanchez:2010af, Lees:2012vv} and Belle \cite{Ha:2010rf}, and the two tagged Belle measurements \cite{Sibidanov:2013rkk}.
The most recent analysis is the Belle  hadronic tagged analysis \cite{Sibidanov:2013rkk},  performed in  2013, which gives
 a branching ratio of
$ {\cal{B}} ( B^0 \to \pi^- l^+ \nu) = (1.49 \pm 0.09_{\mathrm{stat}} \pm 0.07_{syst}) \times 10^{-4}$, whose uncertainty is not very far from
 the more precise
results from untagged measurements.
%
%
By employing
their measured partial branching fractions, and combining    LCSR, lattice points and the BCL \cite{Bourrely:2008za}
parametrization, the
 Belle collaboration extracts the value
$ |V_{ub}|  = (3.52 \pm  0.29) \times 10^{-3}$ \cite{Sibidanov:2013rkk}.

The HFLAV  $|V_{ub}|$  determination comes from a combined fit of a $B \to \pi$ form factor parameterization to theory predictions and the average $q^2$ spectrum in data.
The theory input included in the fit are the results from the FLAG lattice average \cite{Aoki:2016frl} and the light-cone sum rule result at $q^2$ = 0 GeV$^2$ \cite{Bharucha:2012wy}.
For the form factor parametrization,  the BCL parametrization is used \cite{Bourrely:2008za} with 3+1 parameters, i.e. 3 parameters for the coefficients in the BCL expansion and one normalization parameter for $|V_{ub}|$.
The results of the combined fit are \cite{Amhis:2016xyh}
\beq
|V_{ub}| = (3.67 \pm 0.09 \pm 0.12) \times 10^{-3}
\eeq
where the first error comes from the experiment and the second one from theory.

The FLAG Collaboration performs a constrained BCL fit of the vector and scalar form factors, together with the combined
experimental datasets, finding \cite{Aoki:2016frl}
\beq
|V_{ub}| = (3.73 \pm 0.14) \times 10^{-3}
\eeq
The previous $|V_{ub}|$ estimates, together with recent estimates given by
Fermilab/MILC   \cite{Lattice:2015tia} and RBC/UKQCD   \cite{Flynn:2015mha}
Collaborations, have been reported in Table \ref{phidectab03}.

\begin{table}[h]
\centering
\caption{Status of  exclusive $|V_{ub}|$  determinations and indirect fits.}
\label{phidectab03}
\begin{tabular}{lrr}
\hline
{ \color{red}{Exclusive decays}}
& {\color{red}{ $ |V_{ub}| \times  10^{3}$}}
  \\
\hline
{\color{blue}{ $\bar B \rightarrow \pi l \bar \nu_l$  }}    & \\
\hline
HFLAV (FLAG+LCSR, BCL) 2016    \cite{Amhis:2016xyh}  & $3.67 \pm 0.09 \pm 0.12$\\
FLAG 2016    \cite{Aoki:2016frl}  & $3.73\pm 0.14$\\
Fermilab/MILC  2015  \cite{Lattice:2015tia}  & $3.72 \pm 0.16$\\
RBC/UKQCD   2015  \cite{Flynn:2015mha}  & $3.61 \pm 0.32$ \\
\hline
{\color{blue}{ $\bar B \rightarrow \omega l \bar \nu_l$   }}   & \\
\hline
Bharucha et al. 2016 (LCSR)     \cite{Straub:2015ica}  & $3.31 \pm 0.19_{\mathrm{exp}} \pm 0.30_{\mathrm{th}}$\\
\hline
{\color{blue}{ $\bar B \rightarrow \rho l \bar \nu_l$     }}  & \\
\hline
Bharucha et al. 2016 (LCSR)     \cite{Straub:2015ica}   & $3.29 \pm 0.09_{\mathrm{exp}} \pm 0.20_{\mathrm{th}}$\\
\hline
 {\color{blue}{ $ \Lambda_b \rightarrow p \, \mu\nu_\mu$ }}     & \\
\hline
HFLAV  (combined fit excl B)  \cite{Amhis:2016xyh,Fiore:2015cmx}    & $ 3.50 \pm 0.13 $\\
\hline
{\color{blue}{ Indirect fits}} &
\\
\hline
UTfit  (2017) \cite{Utfit16} &
$3.61 \pm  0.12$\\
CKMfitter  (2016, $3 \sigma$) \cite{CKMfitter} &
$ 3.71^{+0.24}_{-0.19}$
\\
\hline
\end{tabular}
\end{table}

Other exclusive meson decays induced by
$b \to u \ell \bar\nu_l$
transitions at
the quark level are   $B \to \rho/\omega \, \ell \bar\nu_l$ decays.
The LCSR computation of the needed form factors  has allowed different estimates of $|V_{ub}|$; recent values have also been reported in Table \ref{phidectab03}.
 Let us observe that the values extracted by $B \to \rho/\omega \, \ell \bar\nu_l$ decays appear to be systematically lower than the ones extracted by  $B \to \pi \ell \nu$ decays.

The $B_s \to K^{(\ast) }\ell \nu$ decays
have not been measured yet; however,  they can become an additional channel to extract $|V_{ub}|$, since they are expected to be within the reach of future $B$-physics facilities  \cite{Flynn:2015mha, Meissner:2013pba, Albertus:2014rna, Bouchard:2014ypa, Feldmann:2015xsa}.

 Another channel depending on  $|V_{ub}|$ is the
baryonic  semileptonic $\Lambda^0_b \to p \mu^- \bar \nu_\mu$ decay. At the end of Run I,
LHCb has measured
 the probability
 of this decay relative to the channel $\Lambda^0_b \to \Lambda^+_c  \mu^- \bar \nu_\mu$ \cite{Aaij:2015bfa}.
This result has been combined with the ratio of form factors computed
using lattice QCD with
2+1
flavors of dynamical domain-wall fermions \cite{Detmold:2015aaa},
 enabling the first determination of the ratio of CKM elements $|V_{ub}|/|V_{cb}|$  from
baryonic decays \cite{Aaij:2015bfa}.
The value of $|V_{ub}|$ depends on the choice of the value of $|V_{cb}|$.
A combined fit from HFLAV for $|V_{ub}|$ and $|V_{ub}|$ that includes the constraint
from LHCb, and the determination of $|V_{ub}|$ and $|V_{ub}|$ from exclusive B meson decays, gives
\cite{Amhis:2016xyh,Fiore:2015cmx}
\beq
|V_{ub}|=( 3.50 \pm 0.13) \times 10^{-3}
\eeq

Indirect determination of $|V_{ub}|$   by the UTfit \cite{Utfit16}  and the CKMfitter \cite{CKMfitter} collaborations
have also been reported in Table
\ref{phidectab03}.

Finally,  let us mention that in 2016 Belle has   presented the first experimental result on $B \to \pi\,  \tau \, \nu$, with an upper limit compatible with the SM \cite{Belle16}.

\section{Inclusive $|V_{ub}|$ determination}

The extraction of $|V_{ub}|$ from inclusive decays requires to address theoretical issues absent in the inclusive $|V_{cb}|$ determination, since the experimental  cuts, needed to reduce the background, enhance the relevance of the so-called threshold region in the phase space. Several theoretical schemes are available, which
are  tailored
to analyze data in the threshold region,  but  differ
in their treatment of perturbative corrections and the
parametrization of non-perturbative effects.
We limit to compare four theoretical different approaches, which have been recently analyzed
by BaBar \cite{Lees:2011fv}, Belle \cite{Urquijo:2009tp}  and  HFAG  \cite{Amhis:2014hma} collaborations, that is: ADFR by Aglietti, Di Lodovico, Ferrera and Ricciardi \cite{Aglietti:2004fz, Aglietti:2006yb,  Aglietti:2007ik}; BLNP
by Bosch, Lange, Neubert and Paz \cite{Lange:2005yw, Bosch:2004th, Bosch:2004cb}; DGE, the dressed gluon exponentiation, by Andersen and Gardi \cite{Andersen:2005mj}; GGOU by Gambino, Giordano, Ossola and Uraltsev \cite{Gambino:2007rp} \footnote{Recently, artificial neural networks have been used to parameterize the shape functions and  extract $|V_{ub}|$ in the GGOU framework \cite{Gambino:2016fdy}. The results are in good agreement with the original paper.}.
Although conceptually quite different, all these approaches
lead to roughly consistent results when the same inputs are used and the
theoretical errors are taken into account.
The HFLAV estimates \cite{Amhis:2016xyh}, together with the latest estimates by BaBar \cite{Lees:2011fv, Beleno:2013jla} and Belle
\cite{Urquijo:2009tp}, are reported in Table \ref{phidectab04}.

\begin{table}[h]
\centering
\caption{Status of inclusive $|V_{ub}|$  determinations.}
\label{phidectab04}
\begin{tabular}{lrrrr}
 \hline
 { \color{red}{ Inclusive decays}} &
& {\color{red}{ $ |V_{ub}| \times  10^{3}$}}
  \\
\hline
& { \color{blue}{ ADFR }}  \cite{Aglietti:2004fz, Aglietti:2006yb,  Aglietti:2007ik}  &  { \color{blue}{  BNLP  }}   \cite{Lange:2005yw, Bosch:2004th, Bosch:2004cb}&   { \color{blue}{  DGE  }}   \cite{Andersen:2005mj} &  { \color{blue}{   GGOU  }}    \cite{Gambino:2007rp} \\
\hline
HFLAV 2016 \cite{Amhis:2016xyh} & $4.08 \pm 0.13^{+ 0.18}_{-0.12}$ & $ 4.44 \pm 0.15^{+0.21}_{-0.22}  $  & $4.52 \pm 0.16^{+ 0.15}_{- 0.16}$ &
$4.52 \pm  0.15^{ + 0.11}_ { -0.14} $  \\
BaBar 2011  \cite{Lees:2011fv} &  $4.29 \pm 0.24^{+0.18}_{-0.19}  $  & $4.28 \pm 0.24^{+0.18}_{-0.20}  $    & $4.40 \pm 0.24^{+0.12}_{-0.13}  $
& $4.35 \pm 0.24^{+0.09}_{-0.10}  $ \\
 Belle 2009 \cite{Urquijo:2009tp} & $4.48 \pm 0.30^{+0.19}_{-0.19}  $ & $ 4.47 \pm 0.27^{+0.19}_{-0.21}  $ &  $4.60 \pm 0.27^{+0.11}_{-0.13}  $ & $4.54 \pm 0.27^{+0.10}_{-0.11}  $ \\
\hline
\end{tabular}
\end{table}
The BaBar and Belle  estimates  in Table \ref{phidectab04} refer to the value extracted by
the
most inclusive measurement, namely the one based on
the two-dimensional fit of the $M_X-q^2$
distribution with
no phase space restrictions, except for
$p^\ast_l > 1.0$  GeV. This selection  allow to access approximately
90\% of the total phase space \cite{Beleno:2013jla}.
The BaBar collaboration also
reports measurements of $|V_{ub}|$
in other regions of the phase space \cite{Lees:2011fv}, but the values reported in  Table \ref{phidectab04} are the most precise.
When averaged, the ADFR value is lower than the one obtained with the other three approaches, and closer to the exclusive values;  this difference
disappears
if we restrict to the BaBar
 and Belle results quoted in  Table \ref{phidectab04}.
By taking the arithmetic average of the
results obtained from these  four different QCD predictions of the partial rate the Babar collaboration gives \cite{Lees:2011fv}
$
|V_{ub}|=(4.33 \pm 0.24_{\mathrm{exp}} \pm 0.15_{\mathrm{th}}) \times 10^{-3}
\label{VinclBabar}
$.
%

By comparing the  results in Table \ref{phidectab03} and \ref{phidectab04}, we observe a tension between exclusive and inclusive determinations, of the order of $2-3\sigma$, according to the chosen values.
Belle II
is expected, at about 50 ab$^{-1}$,  to decrease experimental  errors on both inclusive and exclusive $|V_{ub}|$  determinations up to about 2\% \cite{Lubej:2017wwx}.

A new measurement \cite{TheBABAR:2016lja} from BABAR based on the inclusive electron spectrum
determines the partial branching fraction and $|V_{ub}|$ for $E_e > 0.8$ GeV. This analysis
shows clearly that the partial branching fraction has substantial model dependence when
the kinematic acceptance includes regions dominated by $B  \to  X_c \ell \nu$  background.

\section{Exclusive decays into heavy leptons}

In the SM the  couplings to the $W^\pm$ bosons are assumed to be universal for all leptons. This universality can be tested in
semileptonic $B$ meson decays involving a $\tau$ lepton, which might be sensitive to a possible
charged Higgs boson or other BSM processes.
The ratio of branching fractions (the denominator is the average for $\ell \in \{e, \mu\}$)
\begin{equation}
R_{D^{(\ast)}} \equiv  \frac{{\cal{B}}( B \to D^{(\ast)} \tau \nu_\tau)}{{\cal{B}}( B \to D^{(\ast)} l  \nu_l)}
\label{ratiotau0}
\end{equation}
 is typically used instead of the absolute branching fraction
of $ B \to D^{(\ast)} \tau  \nu_\tau$ decays to cancel  uncertainties common to the numerator and the denominator.
These include the CKM matrix element and several theoretical uncertainties on hadronic form factors and experimental reconstruction effects.

In  the standard model values for $R_{D^\ast}^{SM}$ can be calculated by means of HQE \cite{Fajfer:2012vx}, while the most recent computation of $R_D^{SM}$  uses a fit to lattice and experimental data \cite{Bigi:2016mdz}
\bea
R_{D^\ast}^{SM} &=& 0.252\pm 0.003 \label{FajferRDstar} \\
R_D^{SM}  &=&  0.299 \pm 0.003
\label{ratiotauteo1}
\eea
In  the standard model, estimates by lattice collaborations have become available in 2015 \cite{Na:2015kha, Lattice:2015rga}
\bea
R_{D^\ast}^{HPQCD} &=& 0.300\pm 0.008  \\
R_D^{FL/MLC}  &=&  0.299 \pm 0.011
\label{ratiotauteo2}
\eea
The previous values are all in agreement among them and with
older $ R_{D}^{SM}$ determinations \cite{Kamenik:2008tj, Becirevic:2012jf}.

 Exclusive semi-tauonic $B$ decays were
first observed by the Belle Collaboration in 2007 \cite{Matyja:2007kt}.
Subsequent
analysis by Babar and Belle \cite{Aubert:2007dsa, Bozek:2010xy,Huschle:2015rga} measured
 branching fractions above, although consistent with, the SM predictions.
 In 2012-2013
Babar
 has measured
$R_{D^{(\ast)}}$ by using  its full data sample \cite{Lees:2012xj, Lees:2013uzd},
and reported a significant excess over the SM expectation, confirmed in 2016
 by the first measurement of $R_{D^\ast}$
using the semileptonic tagging method (Belle \cite{Sato:2016svk}).

In 2015 a confirmation came also by the  LHCb collaboration, who has studied the decay $\bar B \to D^{\ast +} \tau \bar \nu_\tau$ with $ D^{\ast +} \to  D^{0}\pi^+$ and $\tau \to \mu \nu_\tau \bar \nu_\mu$ in $pp$ collisions
\cite{Aaij:2015yra}.

Most recently, the Belle collaboration has reported
 a new measurement
 in the hadronic
$\tau$
decay modes
which is statistically independent of the previous Belle
measurements,  with  a  different  background  composition, giving \cite{Hirose:2016wfn}
\begin{equation}
R_{D^\ast} = 0.270\pm 0.035^{+0.028}_{-0.025}
\end{equation}
where the first errors are statistical and the second ones systematic.
This result
is  consistent with the theoretical predictions of the SM in Ref. \cite{Fajfer:2012vx}
 within 0.6$\sigma$ standard deviations.
They also report the first  measurement of the
$\tau$ lepton polarization in the decay $\bar B \to D^\ast \tau^- \bar \nu$  \cite{Hirose:2016wfn}, which is again compatible with SM expectations \cite{Tanaka:2012nw}.

By averaging the most recent measurements  \cite{Lees:2012xj, Lees:2013uzd,Huschle:2015rga,Aaij:2015yra, Sato:2016svk, Hirose:2016wfn}, including results frome LHCb presented at FPCP 2017 \cite{LHCbFPCP2017},  the HFLAV Collaboration has found \cite{HFAG2016}
\bea
R_{D^\ast} &=& 0.304 \pm 0.013 \pm 0.007 \\
R_D  &=& 0.407 \pm 0.039 \pm 0.024  \qquad \qquad
\label{ratiotau}
\eea
where the first uncertainty is statistical and the second one is
systematic. $R_D$ and $R_{D^\ast}$  exceed the SM
 values
  by about  2$\sigma$ and 3$\sigma$, respectively.
If one consider both deviations, the tension rises to about 4$\sigma$.
At Belle II a better understanding of
backgrounds tails under the signal  and a reduction of the uncertainty  to 3\% for $R_{D^\ast}$  and 5\% for   $R_D$ is expected at 5 ab$^{-1}$.

While $R_B$ is defined as the ratio of branching fractions of decays that occur  at tree level in the SM at the lowest perturbative order, the observable  $R_K$ is defined as the ratio of branching fractions of rare decays, starting at one loop order in the SM, that is
\begin{equation}
	R_{K^{(*)}[q^2_{\rm min},q^2_{\rm max}]}=\frac{\mathcal B(B\to K^{(*)} \mu^+\mu^-)_{q^2\in[q^2_{\rm min},q^2_{\rm max}]}}{\mathcal B(B\to K^{(*)} e^+e^-)_{q^2\in[q^2_{\rm min},q^2_{\rm max}]}}
\end{equation}
where $R_{K^{(*)}}$ is measured over specific ranges for the squared di-lepton invariant mass $q^2$ (in GeV$^2$).

 Let us  compare  experimental data and theoretical determinations, and express their tension in terms of $\sigma$
\begin{equation}
\begin{aligned}[l]
	&R_{K[1,6]}^\text{exp}=0.745^{+0.090}_{-0.074}\pm 0.036\text{~\cite{Aaij:2014ora}}\\
	&R_{K^*[0.045,1.1]}^\text{exp}=0.66^{+0.11}_{-0.07}\pm 0.03\text{~\cite{Aaij:2017vbb}}\\
	&R_{K^*[1.1,6.0]}^\text{exp}=0.69^{+0.11}_{-0.07}\pm 0.05\text{~\cite{Aaij:2017vbb}}
\end{aligned}
\quad
\begin{aligned}[l]
	&R_K^\text{th}=1.00\pm 0.01\text{~\cite{Bordone:2016gaq,Capdevila:2017ert}}\\
	&R_{K^*[0.045,1.1]}^\text{th}=0.922\pm 0.022\text{~\cite{Capdevila:2017ert}}\\
	&R_{K^*[1.1,6.0]}^\text{th}=1.000\pm 0.006\text{~\cite{Capdevila:2017ert}}
\end{aligned}
\quad
\begin{aligned}[l]
	& 2.8~ \sigma\\
        & 2.7~ \sigma\\
        & 3.0~ \sigma\\
\end{aligned}
\end{equation}
In the experimental data the first errors are statistical and the second ones systematic.  The impact of radiative corrections  has been estimated not to exceed a few \% \cite{Bordone:2016gaq}.

The alleged   breaking of lepton-flavour universality suggested by most of the data is quite large,  and several theoretical models have been tested against the experimental results.
A welcome feature of measurements in the $\tau$ sector is the capacity of putting  stringent limits on new physics models (see e.g. \cite{Descotes-Genon:2017ptp, Celis:2012dk, Faroughy:2016osc, Becirevic:2016hea,Becirevic:2016yqi,Bordone:2016tex, Crivellin:2016ejn}). In particular,  the simultaneous interpretation of the deviation of $R_D $ and $R_D^\star $ in terms of the two Higgs doublet model II (2HDMII) seems to be ruled out \cite{Lees:2012xj}.
This is also particularly interesting since this
corresponds to the Higgs sector of commonly used supersymmetric models.

\section*{Acknowledgements}

 This work received partial financial support  from MIUR under Project No. 2015P5SBHT and from the INFN research initiative ENP.

\bibliography{VxbRef}

\end{document}